\documentclass[12pt]{amsart}

\newcommand{\HH}{\mathcal H}
\newcommand{\R}{\mathbb R}
\newcommand{\eqdef}{\stackrel{\rm def}{=}}

\begin{document}

\title[Analogs of the Hamilton--Jacobi and Schrodinger equations]{Analogs of the
Hamilton--Jacobi and Schrodinger equations for multidimensional
variational problems of field theory}
\author{A.~V.~Stoyanovsky}

\email{stoyan@mccme.ru}
\address{Moscow Center for Continuous Mathematical Education,
Bolshoj Vlasjevskij per., 11, Moscow, 119002, Russia}
\maketitle
\begin{abstract}
Generalizations of the Hamilton--Jacobi and Schrodinger equations for
multidimensional variational problems of field theory
are deduced. These generalizations are so-called variational differential
equations.
\end{abstract}

In this note we deduce certain analogs of the Hamilton--Jacobi and Schrodinger
equations for theories described by Lagrangians with several independent
variables. For brevity and geometric clearness, below we consider only the case
of two independent variables and one dependent variable, i.e., the action
functional of the form
\begin{equation}
J=\int\!\int F(x,y,z,z_x,z_y)\,dxdy.
\end{equation}
Regarding generalization to an arbitrary number of independent and dependent
variables, see Remark at the end of Subsection 2.

\subsection*{1. Geometric optics in the space of curves}
Let us argue by analogy with geometric optics. The role of the inhomogeneous
medium in which light propagates will be played by the space whose points
are parameterized curves
$C=(x(\tau),y(\tau),z(\tau))$ in $\R^3$. Assume that for each point $C$
and for each infinitely small increment
($\delta x(\tau)$, $\delta y(\tau)$, $\delta z(\tau)$),
the time $\delta T$ of propagation of light from the point $C$ to the point
$C'=(x(\tau)+\delta x(\tau)$, $y(\tau)+\delta y(\tau)$, $z(\tau)+\delta z(\tau))$
is given by the formula
$$
\begin{aligned}
\delta T&=\Phi(x,y,z,\delta x,\delta y,\delta z)\\
&\eqdef \int (\dot y\delta x-\dot x\delta y)
F\left(x,y,z,\frac{\dot y\delta z-\dot z\delta y}{\dot y\delta
x-\dot x\delta y},\frac{\dot z\delta x-\dot x\delta z}{\dot
y\delta
x-\dot x\delta y}\right)\,d\tau,
\end{aligned}
$$
where
$\dot x=\frac{dx}{d\tau}$, $\dot y=\frac{dy}{d\tau}$, $\dot z=\frac{dz}{d\tau}$.
(The reason for this choice of $\Phi$ will become clear below.)
Let us call by the indicatrice the hypersurface
$\Phi(x$, $y$, $z$, $\delta x$, $\delta y$, $\delta z)=1$
in the space of functions $(\delta x(\tau)$, $\delta y(\tau)$, $\delta z(\tau))$
for fixed $x(\tau)$, $y(\tau)$, $z(\tau)$.
The time of propagation of light along the trajectory
$C(\alpha)$, $0\le\alpha\le A$, equals
$$
T=\int_0^A\Phi(x(\tau,\alpha),y(\tau,\alpha),z(\tau,\alpha),
\delta x,\delta y,\delta z),
$$
where $\delta x=\frac{\partial x}{\partial\alpha}d\alpha$,
$\delta y=\frac{\partial y}{\partial\alpha}d\alpha$,
$\delta z=\frac{\partial z}{\partial\alpha}d\alpha$.
It is easy to see that the definition of the functional $\Phi$
is chosen so that the time $T$ coincides with the integral (1) over the surface
covered by the curve $C(\alpha)$. Hence a light ray, i.e., a trajectory
of propagation of light with minimal time, realizes the extreme value of the
integral (1). Here the following difference with the usual geometric optics
occurs: one has not one but infinitely many rays passing through two
given points. These rays correspond to various parameterizations of the
solution surface of the variational problem.

Denote by $S(C)$ the function (called the action or the eikonal)
equal to the minimal time of propagation of light from a fixed point $C_0$
to the point $C$. Let us call by the wave fronts $W(T)$ emitted by the point $C_0$
at the time $T$ the hypersurfaces of the same value $T$ of the function $S$.
Assume that the Huyghens principle holds: the wave front $W(T+\Delta T)$
is the common tangent
hypersurface to the wave fronts emitted by the points of the wave front
$W(T)$ at the time $\Delta T$. The infinitesimal version of this principle,
obtained by letting $\Delta T$ tend to zero, would be the following: the tangent hypersurface
to a wave front at a given point is parallel to the tangent
hypersurface to the indicatrice at the tangent vector to the ray. This implies the
equalities:
\begin{equation}
\frac{\delta S}{\delta x(\tau)}=c\frac{\delta\Phi}{\delta(\delta
x(\tau))},\ \frac{\delta S}{\delta
y(\tau)}=c\frac{\delta\Phi}{\delta(\delta y(\tau))},\ \frac{\delta
S}{\delta z(\tau)}=c\frac{\delta\Phi}{\delta(\delta z(\tau))}.
\end{equation}
Here $c$ is a constant independent of $\tau$; in the right-hand sides
one must substitute, instead of
$\delta x(\tau),\delta y(\tau),\delta z(\tau)$, the tangent vector to the ray,
i.e., to the solution of the variational problem; $\frac{\delta S}{\delta x(\tau)}$, etc.
denote the variational derivatives.
The constant $c$ equals one. This can be seen by multiplying both parts of (2)
by respectively
$\delta x,\delta y,\delta z$, summing up together and integrating over $\tau$.
In both parts of the equality one obtains $\delta S=\Phi$.

\subsection*{2. Analogs of the Hamilton--Jacobi and Schrodinger equations}
Calculating the right-hand sides of (2) and substituting
$\frac{\dot y\delta z-\dot z\delta y}{\dot y\delta
x-\dot x\delta y}=z_x$, $\frac{\dot z\delta x-\dot x\delta
z}{\dot y\delta x-\dot x\delta y}=z_y$,
$\dot z=\dot xz_x+\dot yz_y$, one obtains
\begin{equation}
\begin{aligned}
\frac{\delta S}{\delta x}&=\dot y(F-z_xF_{z_x})-\dot
x(-z_xF_{z_y});\\
\frac{\delta S}{\delta y}&=\dot y(-z_yF_{z_x})-\dot
x(F-z_yF_{z_y});\\
\frac{\delta S}{\delta z}&=\dot yF_{z_x}-\dot xF_{z_y}.
\end{aligned}
\end{equation}
Note that the right-hand sides of (2) depend on the functions
$\delta x$, $\delta y$, $\delta z$ (for which we have a large freedom of choice)
only through the numbers $z_x$ and $z_y$ characterizing the tangent
plane at the point $(x,y,z)$ to the solution surface of the variational problem.

In the formulas (3) for $\frac{\delta S}{\delta x}$ and $\frac{\delta
S}{\delta y}$, the coefficients of $-\dot x$ and $\dot y$ equal to
the components of the energy-momentum tensor with the minus sign.
Formulas (3) agree with the well-known formula for the variation
of the integral (1) under variation of the integration surface
with the moving boundary (see, for example, [1], \S33, formula (17)).
Indeed, this formula has the following form.
Assume that the integration surface $D$ with boundary $\partial D$ in $\R^3$,
which is a solution to the variational problem, varies together with
its boundary. Then
$$
\delta J=\int_{\partial D}\left(\frac{\delta S}{\delta x}\delta
x+\frac{\delta S}{\delta y}\delta y+\frac{\delta S}{\delta
z}\delta z\right)\,d\tau,
$$
where $\frac{\delta S}{\delta x}$, $\frac{\delta S}{\delta y}$, and
$\frac{\delta S}{\delta z}$ are given by (3).

Let us eliminate $z_x$ and $z_y$ from (3) using the relation
$\dot xz_x+\dot yz_y=\dot z$. We obtain two equations for
$\frac{\delta S}{\delta x}$, $\frac{\delta S}{\delta
y}$, and $\frac{\delta S}{\delta z}$. One of them is the
obvious equation $\dot x\frac{\delta S}{\delta x}+\dot y\frac{\delta
S}{\delta y}+\dot z\frac{\delta S}{\delta z}=0$ corresponding to the independence
of the choice of parameterization of the curve. The other equation
$\HH\left(x,y,z,\dot x,\dot y,\dot z,\frac{\delta S}{\delta
x},\frac{\delta S}{\delta y},\frac{\delta S}{\delta z}\right)=0$
(in general nonlinear first order {\it variational differential equation})
can be naturally called the analog of the
Hamilton--Jacobi equation. The question arises whether
one can develop here an analog of the canonical formalism.

\medskip
{\bf Example} (the scalar field in two-dimensional spacetime).
Let $F(x$, $y$, $z,z_x,z_y)=\frac{1}{2}(z_x^2-z_y^2)+p(z)$, where
$p(z)=-\frac{m^2}{2}z^2+\ldots$ is a polynomial in $z$. A calculation gives
the following analog of the Hamilton--Jacobi equation:
\begin{equation}
\frac{1}{2}\left(\left(\frac{\delta S}{\delta z}\right)^2+\dot
z^2\right)+\left(\dot x^2-\dot y^2\right)p(z)+\dot x\frac{\delta
S}{\delta y}+\dot y\frac{\delta S}{\delta x}=0.
\end{equation}

Let us obtain from this equation an analog of the Schrodinger equation
by the usual scheme, i.e., replacing $\frac{\delta S}{\delta x(\tau)}$, $\frac{\delta
S}{\delta y(\tau)}$, $\frac{\delta S}{\delta z(\tau)}$ by respectively
$-i\frac{\delta}{\delta x(\tau)}$, $-i\frac{\delta}{\delta
y(\tau)}$, $-i\frac{\delta}{\delta z(\tau)}$ ($i$ is the square root of $-1$).
One obtains the following linear second order variational differential equation:
\begin{equation}
\begin{aligned}
\frac{1}{2}\left(-\frac{\delta^2\Psi}{\delta z(\tau)^2}+\dot
z^2\Psi\right)&+\left(\dot x^2-\dot y^2\right)p(z)\Psi-i\left(\dot
x\frac{\delta\Psi}{\delta y(\tau)}+\dot y\frac{\delta\Psi}{\delta
x(\tau)}\right)=0;\\
\dot x&\frac{\delta\Psi}{\delta x(\tau)}+\dot
y\frac{\delta\Psi}{\delta y(\tau)}+\dot z\frac{\delta\Psi}{\delta
z(\tau)}=0.
\end{aligned}
\end{equation}
Here $\Psi$ is a function on the space of curves
$(x(\tau),y(\tau),z(\tau))$.

\medskip
{\bf Remark.} The above argument is directly generalized to the case of an arbitrary
number of dependent and independent variables. Instead of
parameterized curves in $\R^3$ one considers parameterized $(m-1)$-dimensional
submanifolds in $\R^{m+n}$, where $m$, $n$ are respectively the numbers of
independent and dependent variables.

A theory of the generalized Hamilton--Jacobi equation is outlined in the book [2].
A development of this note is given in the papers [3,4].

\end{document}